

\input harvmac

\overfullrule=0pt

\def\r{\rho}
\def\a{\alpha}

\def\b{\beta}

\def\d{\delta}

\def\e{\epsilon}

\def\th{\theta}

\def\m{\mu}
\def\n{\nu}
\def\o{\omega}
\def\l{\lambda}

\def\no{\noindent}

\def\qq{\qquad}



\rightline{USC-91/HEP-B6}
\rightline{November 1991}

\bigskip\bigskip

\centerline    {\bf A SUPERSTRING THEORY }
 \centerline  {  {\bf IN FOUR CURVED SPACE-TIME DIMENSIONS   }
{\footnote {$^*$} {Research supported in part by the
U.S. Department  of Energy, under Grant No.
DE-FG03-84ER-40168 }}}


\vskip 1.00 true cm

\centerline { I. BARS and K. SFETSOS}

\bigskip

\centerline {Physics Department}
\centerline {University of Southern California}
\centerline {Los Angeles, CA 90089-0484, USA}

\vskip 1.50 true cm
\centerline{ABSTRACT}
\bigskip

Neveu-Schwarz-Ramond type heterotic and type-II superstrings in four
dimensional curved space-time are constructed as exact $N=1$ superconformal
theories. The tachyon is eliminated with a GSO projection.
The theory is based on the N=1 superconformal gauged WZW model for the anti-de
Sitter coset $SO(3,2)/SO(3,1)$ with integer central extension $k=5$. The model
has dynamical duality properties in its space-time metric that are similar to
the large-small ($R\rightarrow 1/R$) duality of tori. To first order in a
$1/k$ expansion we give expressions for the metric, the dilaton, the Ricci
tensor and their dual generalizations. The curvature scalar has several
singularities at various locations in the 4-dimensional manifold. This
provides a new singular solution to Einstein's equations in the presence of
matter in four dimensions. A non-trivial path integral measure which we
conjectured in previous work for gauged WZW models is verified.

\vfill\eject


Sometime ago exact conformal theories based on the anti-de Sitter (ADS) cosets
$SO(d-1,2)_{-k}/SO(d-1,1)_{-k}$ were introduced as models for
strings propagating in curved space-time in $d=2,3,\cdots ,26$ dimensions
\ref\BN{I. Bars and D. Nemeschansky, Nucl. Phys. B348 (1991) 89.}.
The $N=1$ superconformal models that generalize these were also expressed as
Kazama-Suzuki cosets of the form $SO(d-1,2)_{-k}\times SO(d-1,1)_1/SO(d-
1,1)_{-k+1}$ for $d=2,3,\cdots ,15$ . The $d=2$ bosonic model was interpreted
as a string propagating on the manifold of a two dimensional black hole
 \ref\WIT{E. Witten, Phys. Rev. D44 (1991) 314.}.
Similarly, in three dimensions the bosonic model describes the ADS string
propagating in a singular and more complicated manifold which is a new
solution to Einstein's equations with matter
 \ref\CRE{ M. Crescimanno, ``Geometry and Duality of a Non-Abelian Coset
Model", LBL-30947.}
 \ref\BS{I. Bars and K. Sfetsos, ``Generalized Duality and
 Singular Strings in Higher Dimensions'', USC-91/HEP-B5. Accessible as
\#911054 at hepth@lanl.gov. }
 \ref\FL{E. S. Fradkin and V. Ya. Linetsky, ``On Space-Time Interpretation
of the Coset models in $D<26$ String Theory'', HUTP-91/A044.} .
These manifolds have the interesting property of duality which signals a
shortest distance in string theory. The string geodesics, which correspond to
the classical solutions of the gauged WZW model, have been obtained in \BS\
for any dimension, including $d=4$.

In this paper we will investigate the most interesting case. Namely, the
supersymmetric ADS string in four dimensions and $k=5$ which in a certain
sense is unique. We construct heterotic and type-II string theories based on
this coset. The action for our model, in the conformal gauge, has four parts
$S=S_0+S_1+S_2+S_3$ with

 \eqn\action{ \eqalign {
 &S_0(g)={k\over 8\pi}\int_M d^2\sigma\ Tr(g^{-1}\partial_+g\
g^{-1}\partial_-g)
 -{k\over 24\pi}\int_B Tr(g^{-1}dg\ g^{-1}dg\ g^{-1}dg) \cr
 &S_1(g,A)=-{k\over 4\pi}\int_M d^2\sigma\ Tr(A_-\partial_+gg^{-1}
-\tilde A_+g^{-1}\partial_-g + A_-g \tilde A_+g^{-1}-A_-A_+)\cr
 &S_2(\psi_+ ,A_-)=-{k\over 4\pi}\int_M d^2\sigma\ \psi_+^\mu (iD_-
\psi_+)^\nu  \eta_{\mu\nu}, \qquad S_3(\chi_-)= {k\over 4\pi}\int_M d^2\sigma\
\sum_{a=1}^{22}\ \chi_-^a i\partial_+\chi_-^a
 }}
In addition, there are ghost actions $S_4(b_{L},c_{L},\beta_{L},\gamma_L)$
for left movers and $S_5(b_R,c_R)$ for right movers that are added due to the
superconformal or conformal gauge fixing respectively. This action has $(1,0)$
superconformal symmetry (see below) and is appropriate for the heterotic
string. The type-II string requires $(1,1)$ superconformal symmetry. Its
action follows if $\chi_-^a$ is removed and replaced by $\psi_-^\mu$ that
appears with a gauge covariant kinetic term just like $\psi_+^\mu$. Then
$S_3,\ S_5$ are replaced by $S_3(\psi_-,A_+)$ and
$S_5(b_R,c_R,\beta_R,\gamma_R)$.

In the above, $S_0$ is the global WZW model
 \ref\WITT{E. Witten, Comm. Math. Phys. 92 (1984) 455. }
with $g(\sigma^+,\sigma^-)\in SO(3,2)$. By itself this piece has
$SO(3,2)_L\times SO(3,2)_R$ symmetry. Since $SO(3,2)$ has a non-Abelian
compact subgroup $SO(3)$ the quantum path integral could be defined uniquely
only for $k=integer$ (this was not a restriction for $d=2,3$).
\foot{ The easiest way to see this point is to write $g$ in parametric form
$g=abc$ with $a\in SO(3)\ , b\in SO(2)$ and $c\in SO(3,2)/SO(3)\times SO(2)$
and apply the Polyakov-Wiegman formula
\ref\PW{A. M. Polyakov and P. W. Wiegman, Phys. Lett. B131 (1983) 121. }.
Then $S_0(g)$ decomposes into several pieces one of which is $S_0(a)$ that
can be defined only for integer $k$ since $SO(3)$ is compact \WITT . The
remaining pieces do not present a problem. }
Indeed, we take $k=5$ which is the value required by the {\it total} Virasoro
central charge for the supersymmetric left movers \BN

\eqn\cc{ c_L={3kd\over 2(k-d+1)}=15 \qquad {\rm for} \qquad d=4,\quad k=5 . }
$c_L$ is cancelled by the super ghost system of $S_4$. It is seen that $d=4$ is
unique in the sense that, for $d<10$, it is the only case with integer $k$
solution. The other acceptable integer $k$ solution occurs for $d=10$
($k=\infty )$ which is the original Neveu-Schwarz-Ramond flat string. \foot
{There are also negative integer $k$ solutions for $d=11, 12, 13, 15, 19, 20,
25, 28, 40, 55$, $100,\ \infty $. To insure a single time coordinate one must
take now the coset $SO(4,1)/SO(3,1)$ since $k$ is negative
 \ref\IBCS{I. Bars, ``Curved Space-Time Strings and Black Holes", USC-91/HEP-
B4.}.
However, one expects problems with ghosts etc. for superstring
theories with $d>10$. Of course, $k$ can also be taken integer for other
$d<10$ only if one admits direct products of conformal theories.}
For type-II the central charge of the supersymmetric right-movers is also
$c_R=15$. However, for the heterotic string the bosonic part $SO(3,2)_{-
k}/SO(3,1)_{-k}$ gives

\eqn\cright{ c_R(bose)={10 k\over k-3}-{6k\over k-2} = 15  }
for $k=5$ (already fixed in the action). Since the ghosts in $S_5(b_R,c_R)$
contribute $-26$ we require a $c_R(\chi)=11$ contribution from the free
fermions $\chi_-^a$. Therefore the action $S_3$ contains $22$ free fermions.
This action could be viewed as giving rise to $SO(22)_1$ current algebra
theory for right movers. There are many other ways of obtaining $c_R=11$ as
exact conformal theories based on current algebras.
\foot{Some examples are $[(E_8)_1\times SU(4)_1]$, $[(E_7)_1\times SU(5)_1]$,
$[(E_7)_1\times SU(3)_1\times SU(2)_1\times U(1)]$, $[(E_6)_1\times
SO(10)_1]$, $[(E_6)_1\times SU(4)_2]$, $[SO(10)_2\times SU(3)_1]$, etc. }
Perhaps the most interesting one is $E_7\times SU(3)\times SU(2)\times U(1)$
(all at level $k=1$) since it contains just the gauge group of the Standard
Model and a ``hidden" $E_7$.

The second piece in the action $S_1$ gauges
 \ref\WZW{E. Witten, Nucl. Phys. B223 (1983) 422.
 \semi K. Bardakci, E. Rabinovici and B. Saering, Nucl. Phys. B301
(1988) 151.
 \semi K. Gawedzki and A. Kupiainen, Nucl. Phys. B320 (1989) 625.
 \semi H.J. Schnitzer, Nucl. Phys. B324 (1989) 412.
  \semi D. Karabali, Q-Han Park, H.J. Schnitzer and Z. Yang, Phys. Lett. B216
(1989) 307.
 \semi D. Karabali and H.J. Schnitzer, Nucl. Phys. B329 (1990) 649.}
the Lorentz subgroup $H=SO(3,1)$ which is embedded in $SO(3,2)_L\times
SO(3,2)_R$ with a deformation. As explained in \BS\ the action of the gauge
group could be deformed on the left or the right of the group element $g$. If
the matrix representation of the gauged Lorentz algebra on the left is $t_a$
and the one on the right is $\tilde t_a$ then gauge invariance is satisfied by
$\tilde t_a=g_0^{-1}t_ag_0$ or $\tilde t_a=g_0^{-1}(-t_a)^Tg_0$, where $g_0$
is any constant group element in {\it complexified} $SO(3,2)$ (including
$g_0$'s not continously connected to the identity) and $t^T$ is the transpose
of the matrix. In this notation the action $S_1$ is expressed in terms of
$A_{\pm}=A_{\pm}^at_a$ and $\tilde A_{\pm}= A_{\pm}^a\tilde t_a$ with the same
$SO(3,1)$ gauge potential $A_\pm^a(\sigma^+,\sigma^-)$. The simplest case of
$\tilde t_a=t_a$ corresponds to the standard vector subgroup. The remaining
cases generalize the vector/axial gauging options that were first noticed for
the 2d black hole
\ref\GIV{A. Giveon, ``Target Space Duality and Stringy Black Holes", LBL-30671}
 \ref\IBBH{I. Bars,``String Propagation on Black Holes'', USC-91/HEP-B3.}
 \ref\DVV{R. Dijkgraaf, H. Verlinde and E. Verlinde, ``String Propagation in a
Black Hole Background", PUPT-1252 (1991)}
 \ref\KIR{E. Kiritsis, ``Duality in Gauged WZW Models'', LBL-30747.}
and thus provide a generalization of the concept of duality.
Examples are given in \BS .

The action $S_2$ contains the fermions $\psi_+^\mu $ with $\mu=0,1,2,3$ that
belongs to the coset $SO(3,2)/SO(3,1)$. The flat Minkowski metric
$\eta_{\mu\nu}=diag\ (1,-1,-1,-1)$ is used to contract the Lorentz indices. As
shown in
 \ref\KS{Y. Kazama and H. Suzuki, Nucl. Phys. B321 (1989) 232. }\
coset fermions lead to $N=1$ superconformal symmetry. Indeed the super coset
scheme $SO(3,2)_{-5}\times SO(3,1)_1/SO(3,1)_{-4}$ for left movers
requires that they appear with gauge covariant derivatives $D_-
\psi_+^\mu=\partial_-\psi_+^\mu-(A_-)^\mu{}_\nu\psi_+^\nu$. The explicit
supersymmetry transformations are written more conveniently in terms of the
$5\times 5$ matrix $\psi_+=\left (\matrix {0 & -\psi_+^\nu \cr \psi_{+\mu} &
0\cr }\right )$ that belongs to the $G/H$ part of the Lie algebra
\ref\WITKS{ E. Witten, ``The N Matrix Model and Gauged WZW Models", IASSNS-
HEP-91/26 .}.
\eqn\susy { \delta g=i\epsilon_-\psi_+g , \qquad
 \delta\psi_+=\epsilon_- (gD_+g^{-1})_{G/H}, \qquad \delta \chi_-^a=0,
\qquad  \delta A_\pm=0 , }
with  $\partial_-\epsilon_-(\sigma^+)=0$. In a type-II theory $\psi_-^\mu$
also mixes under  supersymmetry with the group element $g^{-1}$ with a
transformation similar to the one above. The independent right-moving
supersymmetry parameter in this case is $\epsilon_+(\sigma^-)$.

This theory is supplemented with the original GSO projection
\ref\GSO{F. Gliozzi, J. Scherk and D. Olive, Nucl. Phys. B122 (1977) 253.
 \semi M. B. Green, J. H. Schwarz and E. Witten, {\it Superstring Theory},
Vol.1, p.218, (Cambridge 1987) }
adapted to four dimensions. Namely, we construct the operator $(-1)^F$ with
the same prescriptions as \GSO\ and project onto the states $(-1)^F=1$. Let us
describe the effect on the ground states in the Neveu-Schwarz and Ramond
sectors for left movers. In our coset scheme these are conveniently labelled
by the scalar, vector and the two spinor representations of the fermionic
$SO(3,1)_1$. The GSO projection eliminates the scalar and one of the spinor
representations so that the tachyon is eliminated from the theory. The
remaining vector and Weyl spinor form the representations $({1\over 2},{1\over
2})$ and $({1\over 2},0)$ of the Lorentz group in four dimensions. As is well
known this is a covariant space-time supersymmetric vector multiplet and
therefore signals the possibility of space-time supersymmetry in our heterotic
model. The GSO projections for the type-II theory can be chosen such that the
remaining ground state Weyl spinors for the left movers and right movers have
either the opposite or the same chirality. Accordingly the theory will be
called type-IIA or type-IIB respectively. To see whether these theories are
supersymmetric in {\it curved} space-time the target space supercharges
have to be constructed explicitly by a curved space-time modification of the
analysis of
\ref\banks{T. Banks, L.J. Dixon, D. Friedan and E. Martinec, Nucl. Phys. B299
(1988) 613. }.
We will report on our efforts in this direction elsewhere.

In the remainder of this paper we will determine the sigma model geometry of
the 4d super ADS string by finding the metric $G_{\alpha\beta}$, antisymmetric
tensor $B_{\alpha\beta}$, dilaton $\Phi$ and curvature tensors in lowest order
in an expansion in $1/k$. For either the heterotic or type-II cases
the superconformal symmetry dictates the general form of the sigma model
order by order in $1/k$. The leading forms are given in e.g.
\ref\Howe{ P. S. Howe and G. Sierra, Phys. Lett. B148 (1984) 451.
\semi J. Bagger and E. Witten, Nucl. Phys. B222 (1983) 1. }
\ref\callan{C. G. Callan, D. Friedan, E. J. Martinec and M. Perry, Nucl.
Phys. B262 (1985) 593.}.
The $1/k$ correction must be consistent with the vanishing of the beta
functions. Since the computations of the beta functions to order $1/k$ are
insensitive to the presence of fermions \callan\ one can simplify the
calculation by concentrating on the purely bosonic theory $S_0+S_1$.

To study the four dimensional ADS bosonic string it is convenient to write the
10 parameter $SO(3,2)$ group element in parametric form as the product $g=h t $
where $h$ is in the left subgroup $h\in SO(3,1)_L$ and $t$ is in the coset
$t\in SO(3,2)/SO(3,1)_L$. Furthermore $h$ and $t$ are given by

\eqn\htt { h=\left ( \matrix {1 & 0 \cr 0 & h_\mu^{\ \nu} \cr } \right ) ,
\qquad \qquad t=\left (\matrix {b  & -bX^\nu \cr
    bX_\mu  & (\eta_\mu^{\ \nu} + ab X_\mu X^\nu) \cr } \right ),}

\no
where the 6-parameter $SO(3,1)_L$ Lorentz group element ($\eta h^T\eta = h^{-
1}$) can be written in the form $h_\mu^{\ \nu}=[(1+\alpha)(1-\alpha)^{-
1}]_\mu^{\ \nu}$, with $\alpha_{\mu\nu}=-\alpha_{\nu\mu}$ when both indices
are lowered. Furthermore, to insure that $t$ is a $SO(3,2)$ group element one
can take $b=\epsilon (1+X^2)^{-{1\over 2}},\ a=(1-b^{-1})/X^2$ with either
$\epsilon=1$ or $\epsilon=-1$ (when $\epsilon =-1$ the group element is far
from the identity; this will play a role in duality). Let us
first consider the standard undeformed theory ($t_a=\tilde t_a$) in
which the Lorentz gauge transformation acts as $g'=\Lambda g \Lambda^{-1}$,
which means $X'_\mu=\Lambda_\mu^{\ \nu}X_\nu,\ (\alpha')_\mu{}^\nu=(\Lambda
\alpha \Lambda^{-1})_\mu^{\ \nu}$. Using the 6 parameter gauge freedom in
$\Lambda$ one can ``rotate" the group parameters to the form

\eqn\gaugefix{\eqalign{&\a_{02}=\a_{03}=\a_{12}=\a_{13}=0\cr
                       &X_0=X_2=0.}}
This gauge fixing will introduce
a Faddeev-Popov determinant that modifies the measure in the path
integral. It will be evaluated below. The gauge fixed form for $X^\mu$ is
appropriate for a space-like $X\cdot X<0$.
Since $X^2$ is a Lorentz invariant the cases of $X^2>0,\ X^2<0,\ X^2=0$
have to be gauge fixed separately. Fortulately it is possible
to pass from one sign to the other by analytic continuation (and renaming the
basis $\mu=0,1,2,3$). Therefore one can concentrate on space-like $X^2<0$.
(Similar remarks apply to the invariants $\alpha^{\mu\nu}\alpha_{\mu\nu},\
\epsilon^{\mu\nu\lambda\sigma}\alpha_{\mu\nu}\alpha_{\lambda\sigma}$.) The
remaining parameters can be written more conveniently by making a change of
variables of the form

\eqn\param{\eqalign{&X_1=-tanh(2r)\ cos\th,\qq  X_3=-tanh(2r)\ sin\th\cr
  &\a_{01}=-{sinh(2t)\over {cosh(2t)+\e'}},  \qq \a_{23}=-tan(\phi),}}

\no
where $\e'=\pm 1$ also plays a role in duality (when $\e'=-1$ the subgroup
element $h$ is far from the identity). Thus, $\alpha_{01}= tanh(t)\ {\rm or}\
coth(t)$.  No such signs are necessary in the parametric form of $\alpha_{23}$
since the range of the compact angle $0\le \phi \le 2\pi$ already accounts for
$cot(\phi)=tan(\pi-\phi)$. Because of the change of variables in \param\ the
path integral measure is changed by a Jacobian. It will be taken into account
below.

The gauge fields $(A_\pm)_{\mu\nu}$ can be integrated out. The result is
another determinant that modifies the measure. From prior experience in
$d=2,3$ we have learned that the logarithm of this determinant is the dilaton
field $\Phi=log(det M)+ const.$ \BS\ as will be seen below. Therefore, it is
desirable to obtain this expression, which is gauge invariant prior to the
gauge fixing described above. The quadratic piece in $S_1$ can be manipulated
using the antisymmetry of $(A_\pm)_{\mu\nu}$

\eqn\quadr{ Tr(A_- m A_+m^T-A_-A_+) = Tr(A_-(m+1)A_+(m^T-1))
=(A_-)_{\mu\nu}M^{\mu\nu,\lambda\sigma}(A_+)_{\lambda\sigma}, }
The $4\times 4 $ matrix $m$ is given by
\eqn\mm{m=h(1+abX X^T).}
where $X_\mu$ forms a column matrix and $X^T$ is a row.
Integrating out the $A_{\pm}$ gives $(det(M))^{-1}$.
This determinant may be written as products of the eigenvalues of the matrix
$m$. To see this, go to a basis with diagonal $m=diag\ (m_0,m_1,m_2,m_3)$ and
note that \quadr \ takes the form $\sum_{\mu<\nu}(A_-
)_{\mu\nu}(A_+)_{\mu\nu}(m_\mu m_\nu-1)$. Thus,

\eqn\detM{\eqalign{det(M)&=\prod_{\mu<\nu=0}^3 (m_\mu m_\nu-1)\cr
&=1+b^3-{1\over 2}(1+b)^2\bigl((Trm)^2-Tr(m^2)\bigr)\cr
&+b(1+b)(Trm\ Tr(m^{-1})-1)-bTr(m^2)-b^2 Tr(m^2)^{-1}.}}
where the expression has been rewritten in terms of traces and determinants.
This permits restoring $m$ to its general form in \mm . Furthermore it is
convenient to notice $det(m)= det(h)det(1+abXX^T)= b $. Using \mm\ and
calculating the traces one finds

\eqn\detMf{\eqalign{det(M)&=(1-b^2)\bigl\{(1-b)\big [1-2{(X^ThX)^2\over
X^4}\big ] +{1\over 2}(Trh^2-(Trh)^2)\cr & -(1+b){X^Th^2X\over
X^2}+2Trh{X^ThX\over X^2}\bigr\}.}}

\no
This is the gauge invariant form. The gauge fixed version is
\eqn\detMgf{
 Ce^{\Phi}= det(M)= -2\e\ sinh^2(2r)\ [cosh(2r)-\e ]\ sin^2\th\ cos^2\th\  \big
[cosh(2t)-\e'cos(2\phi)\big ]^2 .  }
As indicated this leads to the identification of the dilaton field $\Phi$ that
solves the conformal invariance conditions as will be seen below.

To obtain the effective action after the integration over $A_\pm$ one solves
the classical equations for $A_{\pm}$ and substitutes the solution in the
action \action. Furthermore, one exponentiates all the factors in the measure
to produce the dilaton. After tedius algebra one obtains the gauge fixed form
of the bosonic effective action
\eqn\effaction{ S_{eff}={k\over 2\pi} \int d^2\sigma  \ \sqrt{h} \
\big [ h^{ij}\ G_{\alpha\beta}\partial_i\phi^\alpha\partial_j\phi^\beta -
{1\over 4k} R^{(2)}\Phi \big ] , \qquad \phi^\alpha=(t,r,\th,\phi),  }
where the target space metric $G_{\alpha\beta}$ is given by the line element
$ds^2=G_{\alpha\beta}d\phi^\alpha d\phi^\beta$ and the antisymmetric tensor
$B_{\a\b}$ is zero with our parametrization.
The metric that emerges is
\eqn\metric{ \eqalign {  & ds^2=dr^2
+\ \rho^{2}\ (- {dt^2\over cos^2\th} + {d\phi^2\over sin^2\th})
+\ {1\over \rho^{2}}\ \big [d\th
+ {tan\th \ sinh(2t)\ dt\ - cot\th\ sin(2\phi)\ d\phi \over cosh(2t)-\e'
 cos(2\phi)\  }  \big]^2 \cr
 & \rho^2(r,\e)={{cosh(2r)+\e\over cosh(2r)-\e}}\ = \ tanh^2r \ {\rm
or } \ coth^2r,\qquad \e ,\e' =\pm 1\quad {\rm dual\ patches }  } }
The effects of $\e,\ \e'$ can be reproduced by a dual theory
defined by a deformation of the gauge group with a constant $g_0$ as argued in
\BS .
The metric can be put in a diagonal form by changing variables

\eqn\metricn{ \eqalign {
 & z=\e' cos^2\th\ cos(2\phi)+sin^2\th\ cosh(2t), \qquad
 T=cosh(2t), \qq w=cos(2\phi)  \cr
 &  ds^2=dr^2+ { dz^2\over 4\rho^2\ (z-\e'w)(T-z) }
+{\rho^{2}\over 4} (T-\e' w) \big [ { dw^2\over (z-\e' w)(1-w^2)}
- {dT^2 \over (T-z)(T^2-1) } \big ] } }
The ranges of the new parameters are

\eqn\ine{ z-\e' w\ge 0, \qquad T-z\ge 0, \qquad
\qquad T\ge 1, \qq |w|\leq 1. }
Next we list the components of the Ricci tensor in the new basis
$\phi^\alpha=(T,r,z,w)$. They will be needed to verify
conformal invariance at one loop.

\eqn\Ricci{\eqalign{&R_{rr}=-4\ {\e\ cosh(2r)+3\over sinh^2(2r)}\cr
       &R_{rz}={2\ \e\over sinh(2r)}\ \bigl[{1\over {T-z}}-{1\over
{z-\e'w}}\bigr]\cr
                    &R_{rT}={2\ \e\over sinh(2r)}\ {1\over {T-z}}\cr
                    &R_{rw}={2\ \e\over sinh(2r)}\ {-\e'\over {z-\e'w}}\cr
&R_{zz}=-{1\over 2}\bigl[{1\over (z-\e' w)^2}+{1\over (T-z)^2}\bigr]
  +{1\over {2\r^2\ (T-z)(z-\e' w)}}\
  \bigl[{z^2-1\over {\r^2\ (T-z)(z-\e' w)}}-1\bigr]\cr
&R_{zT}={1\over 2}{1\over {T-z}}\ \bigl[{1\over {z-\e' w}}-{2\over
{T-z}}\bigr]\cr
&R_{zw}=-{\e'\over 2}{1\over {z-\e' w}}\ \bigl[{2\over {z-\e'w}}-{1\over
{T-z}}\bigr]\cr
&R_{TT}={\e \r^3\over sinh(2r)}\ {T-\e' w\over {(T-z)(T^2-1)}}
+{\r^4\over 2}\ {(T-\e'w)(z-\e' w)\over {(T-z)^2(T^2-1)}}
-{1\over 2}\ {1\over (T-z)^2}\cr
&R_{Tw}={\e'\over 2}{1\over {(T-z)(z-\e' w)}}\cr
&R_{ww}=-{\e \r^3\over sinh(2r)}\ {T-\e'w \over {(z-\e'w)(1-w^2)}}
-{\r^4\over 2}\ {(T-z)(T-\e' w)\over {(z-\e' w)^2(1-w^2)}}
-{1\over 2}\ {1\over (z-\e'w)^2}.}}
The scalar curvature deduced from the above is

\eqn\curvature{R=-{16\ \e\ \over {cosh(2r)-\e}}
        -4\r^2\ \bigl[{T-z\over {z-\e'w}}+{z-\e'w\over {T-z}}\bigr]
 +{4\over \r^2}\ {z^2-1\over {(T-z)(z-\e' w)}}.}
It has singularities at

\eqn\sing{ r=0,\qquad  T-z=0\ \ (\th=\pm {\pi\over 2}),  \qquad
  z-\e' w=0\ \ [(\th=0),\quad (t=0 \ \ {\rm for}\ \ \e'cos2\phi=1)].  }

Next we turn to the question of conformal invariance. At one loop (which
corresponds to the leading order in a $1/k$ expansion) the following equations
must be satisfied \callan\

\eqn\conf{ R_{\a\b}=D_{\a}D_{\b}\Phi \qquad
                   c_L=6+{3\over {4\ k}}(D_\a \Phi D^\a \Phi+D^2\Phi),}
where $6=3d/2$ includes the contributions of the left moving supersymmetric 4
bosons and 4 fermions. (Similarly, for the right movers one can write $c_R$ by
replacing the $6$ by $4+11$.) These equations are in fact satisfied by the
expressions given above for the metric, the dilaton and the Ricci tensor. In
particular the dilaton contribution to the central charge reduces to just a
constant $c_L=6+18/k$. This result coincides with the large $k$ expansion of
\cc\ $ c_L=3d/2 + 3d(d-1)/2(k-d+1) \rightarrow 6+18/k $. Similarly it also
coincides with the large $k$ expansion of $c_R$ given by \cright\ plus the
required 11. This is another non-trivial check of our dilaton. The dilaton
contribution to the central charge $18/k$ is indeed independent from the
fermions that come along with the bosons either as left movers or right
movers.

We believe that our metric is a new singular solution of Einstein's equations
in four dimensions in the presence of matter.
\foot{ The physical metric in the low energy theory is $G_{\mu\nu}^{phys}=
e^{\Phi}G_{\mu\nu}$. The curvature scalar for this metric has singularities of
higher degrees at the same locations as before.}
The manifold comes in many patches. First, there are the patches that follow
from analytic continuation of our variables (in order to cover the
possibilities of space-like and time-like $X^2$ etc.). These analytic
continuations have to be done consistently with maintaining an $SO(3,2)$ group
element up to a renaming of the indices (see e.g. \BS ). In any one of these
analytic continuations there can be only one time coordinate. Furthermore,
there are the patches that follow from duality transformations as already
indicated in the expression of the metric in the form $\e ,\e' =\pm 1$. One
has to keep in mind a similar situation in the 2d black hole and 3d ADS string
in order to appreciate these comments by analogy. It is desirable to find
somewhat more global (or fully global) coordinates that give a better
description of this manifold and consequently lead to a satisfactory
interpretation of what it represents (as it was possible to do for the 2d
black hole). We are currently working on such a project.

We now consider the path integral measure. Let us ignore the fermions. To
begin, one has a measure $(Dg) (DA) F(g)$ where $(Dg)$ is the Haar measure for
the group $G=SO(3,2)$ and $(DA)$ is the measure for the gauge field which one
may define as the naive measure $DA_+DA_-$. The additional factor $F(g)$ which
is gauge invariant under the gauge subgroup $H=SO(3,1)$ is a priori allowed
since the measure is required to be invariant only under $H$ and not under the
global symmetry $G_L\times G_R$. In \BS\ we found that conformal invariance
required a nontrivial factor that took the form $F=exp(-\Phi)/\sqrt{-G}$ in
the bosonic $d=2,3$ theories. We conjectured that this form might be true for
all gauged WZW theories. We now verify that indeed $F$ takes this form in the
4 dimensional theory.

The Haar measure for $g=ht \in SO(3,2)$ is $(Dg)= (Dh)(D^4 X) / (1+X^2)^{5/2} $
where $(D^4X)$ is the naive Minkowski (or Euclidean) measure, and $(Dh)$ is
the measure for the subgroup $SO(3,1)$ which will be computed in terms of the
parameters $\alpha_{\mu\nu}$ introduced above. Consider the line element

\eqn\measure{ Tr(h^{-1}dh)^2=Tr({1\over {1-\a^2}}d\alpha
             {1\over {1-\a^2} } d\alpha) }
which defines a group metric for the parameters $\alpha_{\mu\nu}$. The Haar
measure is the square root of the determinant of this metric. The problem of
computing the determinant is similar to computing $det(M)$ above. Using
similar methods one obtains the Haar measure

\eqn\Haarf{(Dg)={(D^6\a)(D^4 X)\over f(\alpha,X)},\qquad
 f(\alpha,X)=det(1-\a^2)^{3/2}\ (1+X^2)^{5/2} .}

Next consider the Faddeev-Popov determinant for the 6 gauge conditions
\gaugefix . Let these 6 functions be denoted by $f_i(g)$. By applying a 6
parameter infinitesimal gauge transformation
$\Lambda_{\mu\nu}=\eta_{\mu\nu}+\omega_{\mu\nu}$ one obtains $\delta f_i=
\Delta_i^{\ \mu\nu}\omega_{\mu\nu}$. The Faddeev-Popov determinant is the
determinant of the $6\times 6$ matrix $\Delta_i^{\ \mu\nu}$. Using the vector
and tensor transformation properties
 $ \d\a_{\m\n}=\o_\m{}^{\l}\a_{\l\n} +\o_\n{}^{\l}\a_{\m\l},\quad
 \d X_{\m}=\o_\m{}^{\l}X_\l $
one can easily figure out the transformation of the gauge functions $f_i$ and
then evaluate $\Delta_i^{\ \mu\nu}$ on the gauge slice $f_i=0$. The result is

\eqn\transfgf{\pmatrix{\d\a_{02}\cr\d\a_{03}\cr\d\a_{12}\cr\d\a_{13}\cr
\d X_0\cr\d X_2}=\pmatrix{0&0&\a_{23}&\a_{01}&0&0\cr
                          0&-\a_{23}&0&0&\a_{01}&0\cr
                          0&\a_{01}&0&0&\a_{23}&0\cr
                          0&0&\a_{01}&-\a_{23}&0&0\cr
                          -X_1&0&-X_3&0&0&0\cr
                          0&0&0&X_1&0&X_3}
\pmatrix{\o_{01}\cr\o_{02}\cr\o_{03}\cr\o_{12}\cr\o_{13}\cr\o_{23}.}}
Then the Faddeev-Popov determinant is $\Delta= -X_1 X_3
(\a_{23}^2+\a_{01}^2)^2 $.

By changing variables as in \param\ one picks up a Jacobian
$(DX_1)(DX_3)(D\a_{01})(D\a_{23})= (Dt)(Dr)(D\th)(D\phi) \ J(t,r,\th ,\phi)$.
In terms of these parameters $\phi^\alpha=(t,r,\theta,\phi)$ we have

\eqn\vic {\eqalign{
 &\Delta=-tanh^2(2r)\ sin\th\ cos\th\ { (cosh(2t)-\e'cos(2\phi))^2
  \over cos^4(\phi) \ (cosh(2t)+\e')^2  }\cr
  &J= {4\e'\ sinh(2r)\over {cosh^3(2r)\ cos^2(\phi) \ (cosh(2t)+\e')}}\cr
  &f(\a,X)={8\over {cosh^5(2r)\ cos^6(\phi)\  (cosh(2t) + \e')^3}}.}}
\no
The integration measure becomes $(D^4\phi)\ FJ \Delta / (f\ detM)$. It is
interesting to observe that the known factors combine to give $(D^4\phi)\
\sqrt {-G}\ F$
\eqn\detg{ {J\ \Delta\ \over f\ (detM)} =\sqrt {-G}  =
   {\rho(r,\epsilon) \over |cos\th\ sin\th| } .  }
A similar relation holds for the two dimensional black hole and, in the less
trivial case, for the three dimensional ADS string \BS. We are by now convinced
that one can prove a theorem which states that the various measure factors
(except $F$) combine to give $\sqrt {-G}$ in general for any gauged WZW model.
We now require that the
contribution from  all the measure factors to the effective action be just the
correct dilaton which we have already identified. This fixes $F$
\eqn\FFF{F= { f\over J\Delta} ={e^{-\Phi}\over \sqrt{-G}}.}
As pointed out earlier this factor must be consistent with gauge invariance
under $SO(3,1)$ transformations. This is indeed the case since $e^{-\Phi}$ is
written in a manifestly invariant form in \detMf\ and one can also rewrite

\eqn\covg{G^{-1} =  {[1-\e\sqrt{1+X^2}]^2 \over 4X^2} \bigl[1-
{(4X^ThX-Trh)^2\over {2Trh^2-(Trh)^2+8}}\bigr] .            }

Thus, using the invariant forms \detMf\ \covg\ one has the correct measure in
any gauge. It is interesting to speculate that \FFF\ which yields two
expressions for $F$ may be valid to all orders in $1/k$ when the exact dilaton
and metric forms are inserted. It is natural to expect this since the first
expression is purely group theoretical and is independent of $k$, while the
second expression depends on the higher order corrections to the dilaton and
metric. This idea can be tested for the 2d black hole for which the exact
dilaton and metric have been conjectured \DVV . Using the exact expressions
in \DVV\ we find that indeed the measure factor $F$ is independent of $k$ to
{\it all orders } and satisfies the relation \FFF .

We have formulated heterotic and type-II superstrings in curved space-time in
four dimensions, as exact superconformal field theories. The tachyon
instability has been eliminated through a GSO projection. These models are
special since they can be analysed using current algebra techniques. In the
next stage the quantum theory should be investigated in more detail than \BN\
, including a discussion of modular invariance which is guided by the
Lagrangian formulation. It is expected that the deformation $g_0$ that
appears in the action \action\ plays a role in combining left and right movers
in modular invariant physical states. This process is expected to lead to a
determination of the number of families and related quantities which may be of
phenomenogical significance. Our curved space-time theory may be applicable to
the physics of the early Universe.

\listrefs
\end